\documentclass[12pt]{article}
\usepackage{epsfig}
\usepackage{graphicx}
\usepackage{amsmath}
\usepackage{hhline}
\usepackage{amssymb}
\usepackage{times}
\usepackage{cite}

\newlength{\dinwidth}
\newlength{\dinmargin}
\begin{document}  
% The rest
\newcommand{\pom}{{I\!\!P}}
\newcommand{\reg}{{I\!\!R}}
\newcommand{\slowpi}{\pi_{\mathit{slow}}}
\newcommand{\fiidiii}{F_2^{D(3)}}
\newcommand{\fiidiiiarg}{\fiidiii\,(\beta,\,Q^2,\,x)}
\newcommand{\n}{1.19\pm 0.06 (stat.) \pm0.07 (syst.)}
\newcommand{\nz}{1.30\pm 0.08 (stat.)^{+0.08}_{-0.14} (syst.)}
\newcommand{\fiidiiiful}{F_2^{D(4)}\,(\beta,\,Q^2,\,x,\,t)}
\newcommand{\fiipom}{\tilde F_2^D}
\newcommand{\ALPHA}{1.10\pm0.03 (stat.) \pm0.04 (syst.)}
\newcommand{\ALPHAZ}{1.15\pm0.04 (stat.)^{+0.04}_{-0.07} (syst.)}
\newcommand{\fiipomarg}{\fiipom\,(\beta,\,Q^2)}
\newcommand{\pomflux}{f_{\pom / p}}
\newcommand{\nxpom}{1.19\pm 0.06 (stat.) \pm0.07 (syst.)}
\newcommand {\gapprox}
   {\raisebox{-0.7ex}{$\stackrel {\textstyle>}{\sim}$}}
\newcommand {\lapprox}
   {\raisebox{-0.7ex}{$\stackrel {\textstyle<}{\sim}$}}
\def\gsim{\,\lower.25ex\hbox{$\scriptstyle\sim$}\kern-1.30ex%
\raise 0.55ex\hbox{$\scriptstyle >$}\,}
\def\lsim{\,\lower.25ex\hbox{$\scriptstyle\sim$}\kern-1.30ex%
\raise 0.55ex\hbox{$\scriptstyle <$}\,}
\newcommand{\pomfluxarg}{f_{\pom / p}\,(x_\pom)}
\newcommand{\dsf}{\mbox{$F_2^{D(3)}$}}
\newcommand{\dsfva}{\mbox{$F_2^{D(3)}(\beta,Q^2,x_{I\!\!P})$}}
\newcommand{\dsfvb}{\mbox{$F_2^{D(3)}(\beta,Q^2,x)$}}
\newcommand{\dsfpom}{$F_2^{I\!\!P}$}
\newcommand{\gap}{\stackrel{>}{\sim}}
\newcommand{\lap}{\stackrel{<}{\sim}}
\newcommand{\fem}{$F_2^{em}$}
\newcommand{\tsnmp}{$\tilde{\sigma}_{NC}(e^{\mp})$}
\newcommand{\tsnm}{$\tilde{\sigma}_{NC}(e^-)$}
\newcommand{\tsnp}{$\tilde{\sigma}_{NC}(e^+)$}
\newcommand{\st}{$\star$}
\newcommand{\sst}{$\star \star$}
\newcommand{\ssst}{$\star \star \star$}
\newcommand{\sssst}{$\star \star \star \star$}
\newcommand{\tw}{\theta_W}
\newcommand{\sw}{\sin{\theta_W}}
\newcommand{\cw}{\cos{\theta_W}}
\newcommand{\sww}{\sin^2{\theta_W}}
\newcommand{\cww}{\cos^2{\theta_W}}
\newcommand{\trm}{m_{\perp}}
\newcommand{\trp}{p_{\perp}}
\newcommand{\trmm}{m_{\perp}^2}
\newcommand{\trpp}{p_{\perp}^2}
\newcommand{\alp}{\alpha_s}

\newcommand{\alps}{\alpha_s}
\newcommand{\sqrts}{$\sqrt{s}$}
\newcommand{\LO}{$O(\alpha_s^0)$}
\newcommand{\Oa}{$O(\alpha_s)$}
\newcommand{\Oaa}{$O(\alpha_s^2)$}
\newcommand{\PT}{p_{\perp}}
\newcommand{\JPSI}{J/\psi}
\newcommand{\sh}{\hat{s}}
\newcommand{\uh}{\hat{u}}
\newcommand{\MP}{m_{J/\psi}}
\newcommand{\PO}{I\!\!P}
\newcommand{\xbj}{x}
\newcommand{\xpom}{x_{\PO}}
\newcommand{\ttbs}{\char'134}
\newcommand{\xpomlo}{3\times10^{-4}}  
\newcommand{\xpomup}{0.05}  
\newcommand{\dgr}{^\circ}
\newcommand{\pbarnt}{\,\mbox{{\rm pb$^{-1}$}}}
\newcommand{\gev}{\,\mbox{GeV}}
\newcommand{\WBoson}{\mbox{$W$}}
\newcommand{\fbarn}{\,\mbox{{\rm fb}}}
\newcommand{\fbarnt}{\,\mbox{{\rm fb$^{-1}$}}}
%
% Some useful tex commands
%
\newcommand{\qsq}{\ensuremath{Q^2} }
\newcommand{\gevsq}{\ensuremath{\mathrm{GeV}^2} }
\newcommand{\et}{\ensuremath{E_t^*} }
\newcommand{\rap}{\ensuremath{\eta^*} }
\newcommand{\gp}{\ensuremath{\gamma^*}p }
\newcommand{\dsiget}{\ensuremath{{\rm d}\sigma_{ep}/{\rm d}E_t^*} }
\newcommand{\dsigrap}{\ensuremath{{\rm d}\sigma_{ep}/{\rm d}\eta^*} }
\newcommand{\dedx}{\ensuremath{{\rm d} E/{\rm d} x}}
% Journal macro
\def\Journal#1#2#3#4{{#1} {\bf #2} (#3) #4}
\def\NCA{Nuovo Cimento}
\def\RPP{Rep. Prog. Phys.}
\def\ARNPS{Ann. Rev. Nucl. Part. Sci.}
\def\NIM{Nucl. Instrum. Methods}
\def\NIMA{{Nucl. Instrum. Methods} {\bf A}}
\def\NPB{{Nucl. Phys.}   {\bf B}}
\def\NPPS{Nucl. Phys. Proc. Suppl.} 
\def\NPPSC{{Nucl. Phys. Proc. Suppl.} {\bf C}}
\def\PR{Phys. Rev.}
\def\PLB{{Phys. Lett.}   {\bf B}}
\def\PRL{Phys. Rev. Lett.}
\def\PRD{{Phys. Rev.}    {\bf D}}
\def\PRC{{Phys. Rev.}    {\bf C}}
\def\ZPC{{Z. Phys.}      {\bf C}}
\def\EJC{{Eur. Phys. J.} {\bf C}}
\def\EPL{{Eur. Phys. Lett.} {\bf}}
\def\CPC{Comp. Phys. Commun.}
\def\NP{{Nucl. Phys.}}
\def\JPG{{J. Phys.} {\bf G}} 
\def\EPC{{Eur. Phys. J.} {\bf C}}
\def\PRSL{{Proc. Roy. Soc.}} {\bf}
\def\PETF{{Pi'sma. Eksp. Teor. Fiz.}} {\bf}
\def\JETPL{{JETP Lett}}{\bf}
\def\IJTP{Int. J. Theor. Phys.}
\def\HJ{Hadronic J.}

%\noindent
%DESY ??-???  \hspace*{8.5cm} ISSN ????-???? \\
%16 March 2003.
 
%\vspace*{3cm}                

%\vspace*{1cm}
%%%%%%%%%%%% Begin paper %%%%%%%%%%%%%%%%%%%%.

%\begin{titlepage}
\begin{flushleft}
%{\tt \today } \\
%Deadline for comments is 19 Feb 2007.  \\ 
\end{flushleft}
%\vspace*{2.cm}
\begin{center}
\begin{Large}
{\boldmath \bf Solar Activity and the Mean Global Temperature} \\

\end{Large}
 
\begin{flushleft}
 
A.D.~Erlykin$^{1a}$, 
T.~Sloan$^{2}$            
and A.W.~Wolfendale$^{1}$   

\bigskip{\it

 $ ^{1}$ Dept.of Physics, Durham University, UK.  \\

 $ ^{2}$ Dept of Physics, University of Lancaster, UK.  \\

\smallskip
 $ ^a$ Permanent address P.N. Lebedev Institute, Moscow, Russia.  \\}
%\bigskip
%$~$ Work supported by the John Taylor Foundation. \\}
%\input{ACoRNEauts}

\end{flushleft}
\end{center}

%\vspace*{2.5cm}

\begin{abstract}
\noindent
The variation with time from 1956-2002 of the globally averaged rate of ionization 
produced by cosmic rays in the atmosphere is deduced and shown to have a cyclic 
component of period roughly twice the 11 year solar cycle period. Long term 
variations in the global average surface temperature as a function of time 
since 1956 are found to have a similar cyclic component.  The cyclic variations 
are also observed in the solar irradiance and in the mean daily sun spot number. 
The cyclic variation in the cosmic ray rate is observed to be delayed by 2-4 years 
relative to the temperature, the solar irradiance and daily sun spot variations  
suggesting that the origin of the correlation is more likely to be direct solar 
activity than cosmic rays. Assuming that the correlation is 
caused by such solar activity, we deduce that the maximum recent increase in 
the mean surface temperature of the Earth which can be ascribed to this 
activity is $~\lesssim$14$\%$ of the observed global warming.   

\end{abstract}

%\begin{center}
%(Submitted to ?)
%\end{center}

%\end{titlepage}

%\begin{flushleft}
%\input{ACoRNEauts}
%\end{flushleft}

%\newpage
 
%=========================================================================
\section{Introduction}

Ionization of the air occurs due to cosmic rays (CR), from the decay of trace 
radioactive isotopes, ionization by solar ultra violet light and electrical 
effects such as lightning.  At cloud forming altitude ($>$ 1000 m) 
over the land and at all altitudes over the sea CR are thought to 
dominate the production of ionization in the troposphere \cite{What}.  
Recently, detailed computations of the total ionization rates produced by CR 
in the atmosphere have become available \cite{Usoskin,Usoskin2,Usoskin4,Desorgher}, 
including the time variation from 1951-2004 arising from the changing solar activity. 
In addition, long term data on the charged particle fluxes in the atmosphere are now 
available \cite {Baz1,Baz}. 

It was suggested long ago that CR could be connected 
with the weather and the climate \cite{Ney} and various mechanisms have 
been  suggested \cite{Tinsley, Yu, Harrison} (for reviews see 
\cite{PB,Carslaw}).    
Much publicity has been given to the observation that the reduction in 
the low cloud cover (LCC) observed during solar cycle 22 correlates well  
with the decrease in the cosmic ray (CR) rate as measured by 
neutron monitors \cite{Friis,Palle,MandS}. This led the groups to hypothesise 
that the reduction was caused by the influence of ionization from CRs on 
cloud cover. Furthermore, it has been suggested \cite{MandS,MandS2} that 
this is a significant contributor to global warming. The basis of the 
suggestion is that the cosmic ray rate has been observed to decrease over 
the last century \cite{LandS}. This leads to less ionization in the 
atmosphere, reducing cloud cover according to the hypothesis, allowing 
more sunlight to warm the Earth. This suggestion has been 
questioned on the grounds of inconsistencies between different methods of 
measuring cloud cover \cite{KSKK} and on the grounds of imperfect data  
analyses \cite{Laut}. Attempts have been made to look for local or 
regional correlations which find either nothing \cite{Bradley}, the 
opposite correlation \cite{Udel} or some correlation \cite{Harrison,Vieira}.     
We discount these in order to investigate the hypothesis further and 
on a global scale.  We further discount the likelihood that CR effects 
would change mainly the depth of the clouds, rather than the cloud cover. 
The suggestion was also questioned in a study of the long term CR  
rate \cite{LandF} where it was shown that this rate began to 
increase in 1985 yet global warming continued.  
In a previous publication \cite{SandW} we showed that 
less than 23$\%$ of the observed reduction in cloud cover in solar cycle 22, 
at the 95$\%$ confidence level,
can be ascribed to ionization from CR. Nevertheless, there may be some 
connection between clouds and ionization since it is well known that 
charged drops grow at smaller radii than uncharged drops, providing that 
the supersaturation is high enough \cite{Mason}.

In this paper we report a search for such an effect by attempting to correlate  
directly changes in the cosmic ray rate of ionization with changes in the surface 
temperature of the Earth either through clouds or via some other mechanism. 
First a study is made of the variation in the long term rate of ionization 
produced by CR. We then study the variation of the global average surface 
temperature, the mean daily sun spot number and the solar irradiance with time 
and show a correlation with the variation of the cosmic 
ray signal since 1956. From this variation a limit on the overall temperature 
rise over the last half century that can be ascribed to 
variable solar activity is deduced.  

\section{Long term variation of the rate of ionization due to CR}

Computer simulations of the production of ionization by CR in the atmosphere have 
been made using the CORSIKA \cite{CORSIKA} simulation 
program \cite{Usoskin,Usoskin2,Usoskin4} and the GEANT4 \cite{Geant} 
simulation package \cite{Desorgher}. These simulations agreed with the 
available fragmentary ionization data and with each other to 10$\%$ precision 
\cite{Usoskin2,Usoskin4}. Recently independent long term data on the 
CR flux in the atmosphere using balloon borne Geiger counter measurements 
have become available \cite{Baz1,Baz}. 
Information is also available from the cosmogenic nuclei $^{10}$Be and 
$^{14}$C. However, there are many complications in using such nuclei 
as a proxy for the ionization in the atmosphere \cite{Nikitin}.  
The ionization measurements and the simulations both show an 11 year 
modulation due to the effects of the changing solar wind on the CR primaries.  
The amplitude of this modulation varies with the magnetic latitude  
on the Earth due to the geomagnetic field (see \cite{Usoskin3} for a review 
of these effects). The geomagnetic field causes a cut off for low CR 
rigidities. The minimum vertical rigidity primary which can hit the Earth's 
surface defines the vertical rigidity cut off (VRCO) \cite{QARM}.

The simulations and the balloon data are used here to assess the long term variation 
of the ionization rate in the atmosphere due to CR. To average over the periodic 
variations due to the 11 year solar cycle the method adopted by Lockwood and 
Fr\"ohlich \cite{LandF} was used. In this method the average at a particular 
time, t, is taken as the average over the full solar cycle length starting 
one half a cycle earlier than t. The time dependent cycle 
lengths were taken from figures 3(b) and 4(b) in \cite{LandF}. 

\begin{figure}[ht]
\includegraphics[width=40pc]{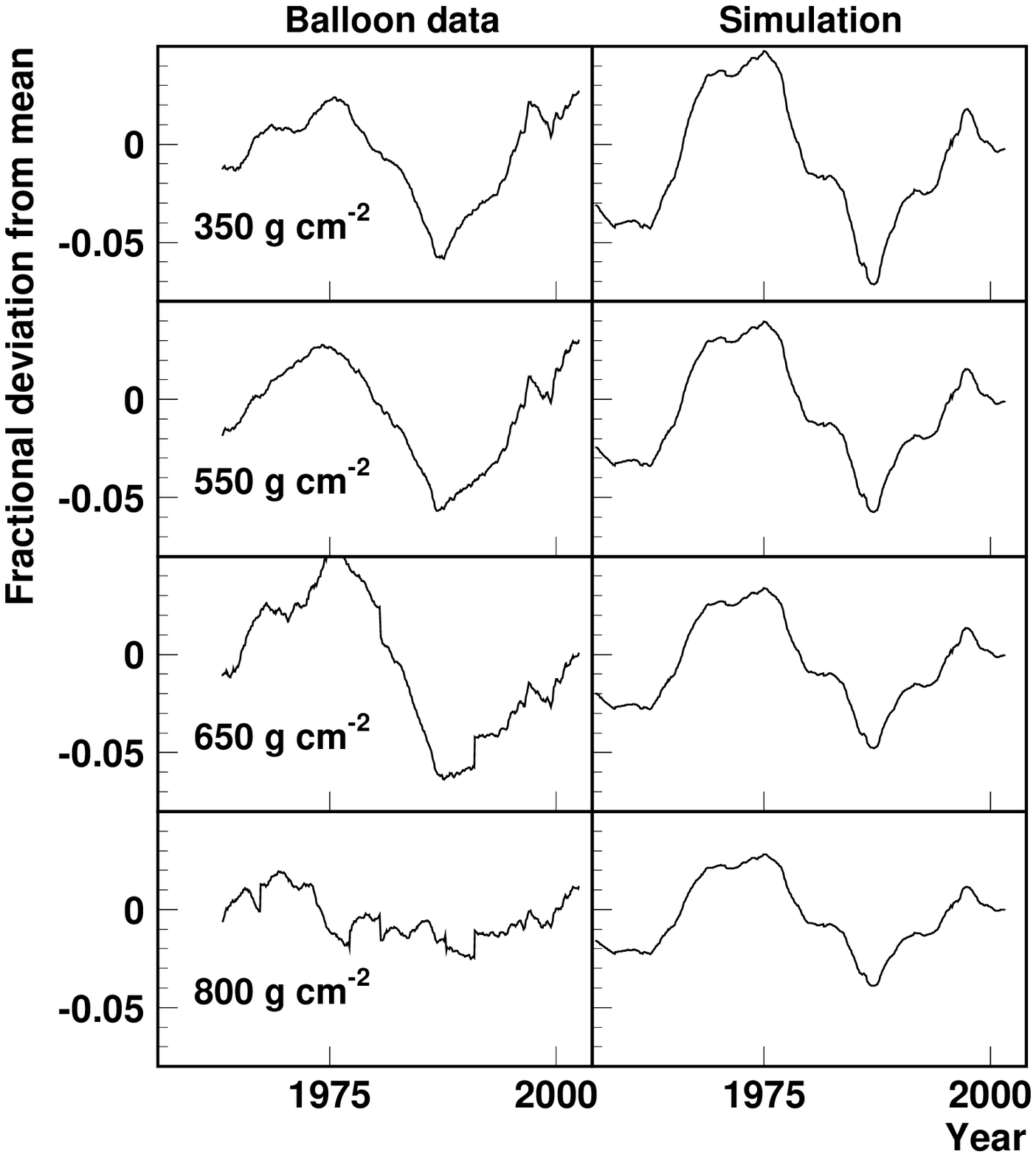}
%\hspace{2pc}%
%\vspace{-20pc}
%\begin{minipage}[b]{18pc}
\caption{\label{fig1}Long term variation of the rate of the flux of charged 
particles in the atmosphere as measured from the Moscow ballon 
data \cite{Baz1,Baz} (the left hand panels). The altitude for each set 
of data is indicated in the  panels. The right hand panels 
show the results of the simulation \cite{Usoskin2} of the ionization rate 
from CR at each altitude.  The smoothing described in the 
text has been applied to each data set to eliminate the effects of the 
11 year solar cycle.} 
%\end{minipage}
\end{figure}

Figure \ref{fig1} shows the count rates from the balloon measurements 
\cite{Baz1,Baz} as a function of time at different altitudes over Moscow with 
the 11 year smoothing applied. There were also measurements over Murmansk 
which gave similar results to the Moscow data and from the 
Antarctic (Mirniy station) which were much more noisy and difficult to 
interpret. The figure also shows the results of the simulations from  
\cite{Usoskin2,Usoskin} at the value of the  VRCO of 2 GV (close to the value 
at Moscow). 

The measurements and the simulation show a similar cyclic behaviour 
with a period of roughly twice the 11 year solar cycle and 
amplitudes in the fractional deviations from the mean 
varying from 3$\%$ at the lowest 
altitude to 5$\%$ at higher altitudes. The simulated 
values of the fractional deviation from the mean agree with the 
measurements to within a factor of two (see figure \ref{fig1}). This 
represents the uncertainty in the procedure to assess the long term 
variation of this quantity. Since the values from the simulation are 
less noisy and are uncontaminated by background they will be used in 
the following as the long term time variation of the rate of ionization 
in the atmosphere by CR.  

The amplitude of the observed cyclic variation of the long term CR rate varies 
with VRCO. The value of the VRCO averaged over the Earth is 8 GV. 
In the remainder of this paper we use the simulation for a VRCO of 8 GV 
at an altitude of 825 g cm$^{-2}$ (2000 m) to represent the change in the 
global mean ionization rate in the atmosphere at cloud forming height. 

Figure \ref{fig2} shows the variation of the simulated ionization rate due 
to CR at a VRCO of 8 GV and altitude of 825 g cm$^{-2}$ as a function of 
time averaged over the 11 year solar cycle. Also shown in figure \ref{fig2} 
is the mean daily sunspot number (SSN) \cite{SSN} and the mean solar irradiance 
(SI) \cite{wang} similarly averaged over the 11 year solar cycle. 
Interestingly, there is a reasonable correspondence between the three 
curves, illustrating the connection between the three phenomena. 
However, the CR changes are delayed by 2-4 years relative to the 
SSN and SI changes. The delay is somewhat longer than those observed between 
the SSN and the CR changes due to the 11 year solar cycle \cite{Usoskin1}. 
This indicates that the long term and shorter term 11 year cycle in the CR 
variation may be influenced by different components of the solar wind.

%The 11 year smoothed $^{10}$Be are also shown in figure \ref{fig1}. They show a 
%different pattern of behaviour from that seen in the balloon measurments, the neutron 
%monitor data and the simulated ionization rates. In addition the amplitude of the 11 year 
%solar modulation is different from that seen in either the neutron monitors or in the 
%simulations of the ionization rate.  The comogenic isotope $^{10}$Be is produced 
%by different types of interactions, by primaries of a different energy and at a 
%different altitude than the secondary particles responsible for the ionization produced by CR. 
%Hence it is not surprising that  $^{10}$Be is not a good proxy for the ionization in 
%the atmosphere.  

\section{Solar modulation of the mean Earth's surface temperature}

\begin{figure}[ht]
\includegraphics[width=33pc]{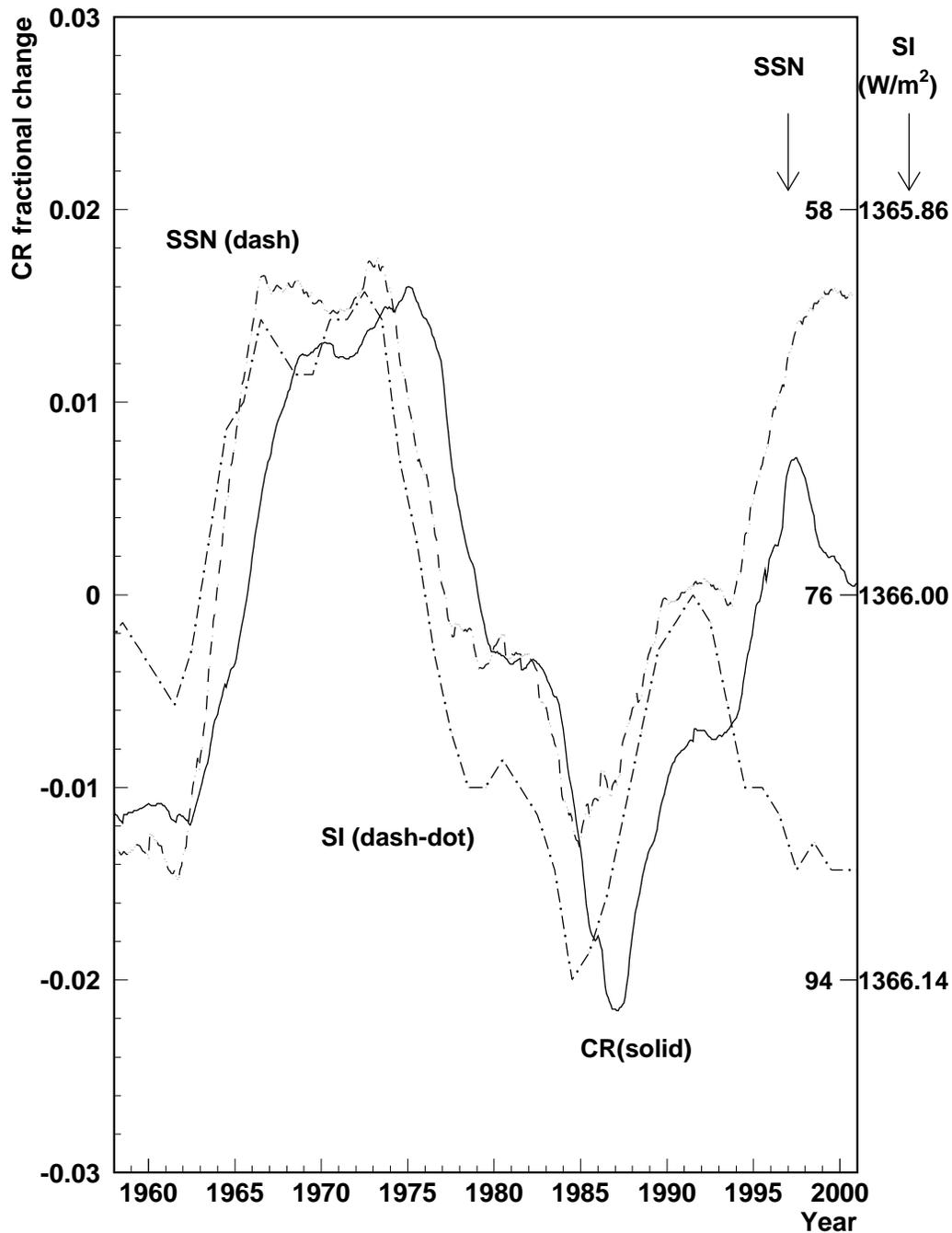}
%\hspace{2pc}%
%\vspace{-20pc}
%\begin{minipage}[b]{18pc}
\caption{\label{fig2} The solid curve shows the fractional change in the CR 
ionization rate from the simulation with time, at VRCO=8GV and altitude 2000m, 
 after the 11 year smoothing 
described in the text. The dashed and dash-dot curves show the mean daily 
SSN and the SI data from \cite{Lean} each with the 11 year smoothing applied. 
The scales for the SSN and SI are shown on the right hand axes, NB these two 
scales are inverted to illustrate the correlation i.e. they increase 
vertically downwards.}
\end{figure}

The data on the global surface temperature are now examined 
to see if a correlation can be found with the changes in solar activity shown 
in figure \ref{fig2}. Such changes could be expected since it is thought that 
the modulation of the global surface temperature due to the 11 year 
solar cycle is approximately 0.1$^\circ$C peak to peak \cite{Loon,Lean}. 

\begin{figure}[ht]
\includegraphics[width=33pc]{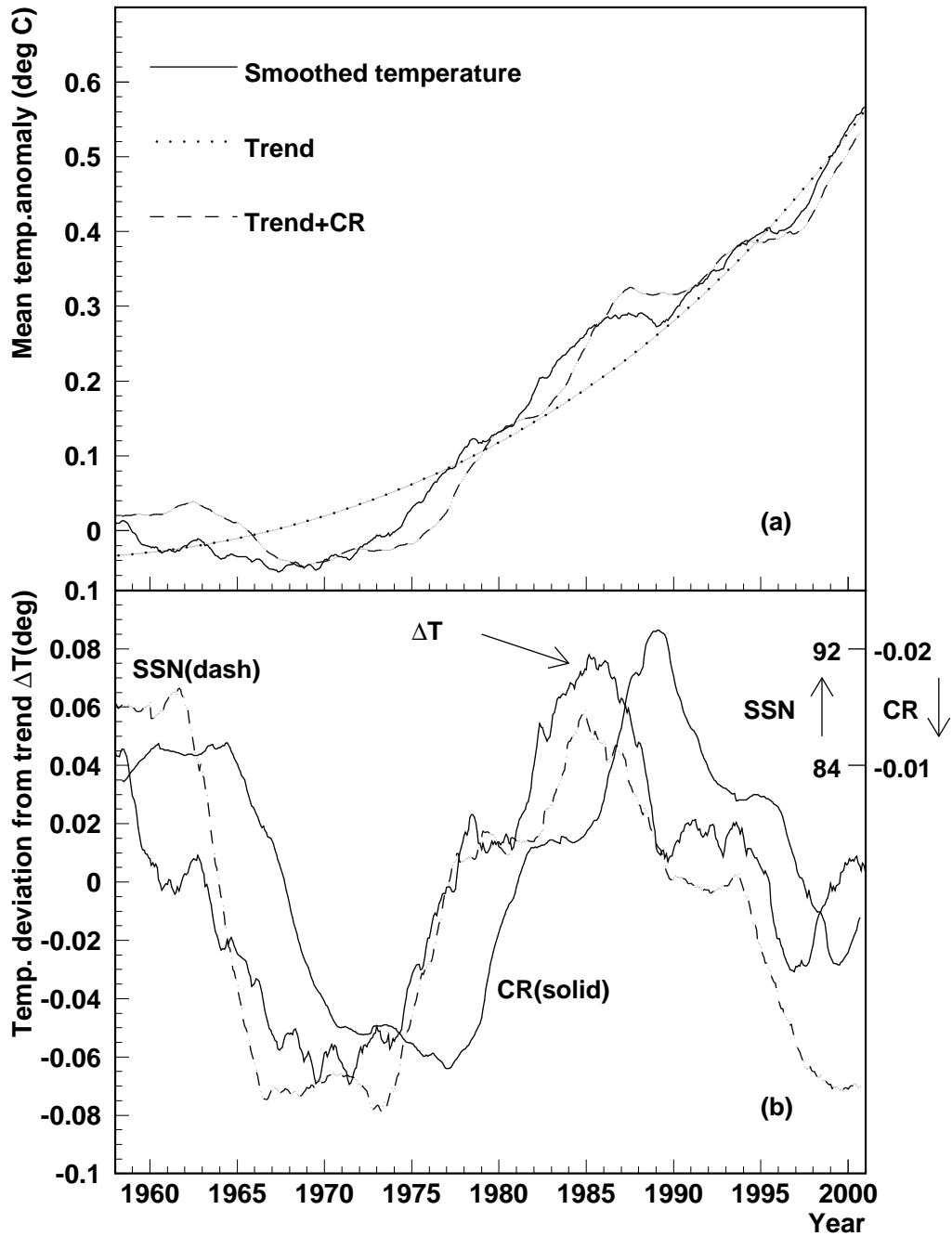}
%\hspace{2pc}%
%\vspace{-20pc}
%\begin{minipage}[b]{18pc}
\caption{\label{fig3}a) - the global surface temperature anomaly (solid curve) \cite{giss}  
with the 11 year smoothing described in the text. The dotted 
curve shows the smooth trend obtained by fitting an empirical function (see text). The dashed 
line shows the long term CR rate from figure \ref{fig2} (inverted and normalised) 
added to the smooth trend curve. (b) The deviations from the trend line in the 
upper panel for the temperature (solid curve labelled $\Delta$T), for the 
CR - labelled CR(solid) - and for the SSN - labelled SSN (dash).  The  scales for  
the latter two are given on the right hand axis.  NB the scale for the CR rate 
is inverted to illustrate the correlation i.e. it increases vertically downwards.}
\end{figure}

Figure \ref{fig3}(a) shows the global surface temperature 
\cite{giss}, as a function of time with the same 11 year smoothing 
as that described above. The data show an oscillatory behaviour about 
a smooth upward trend. The smooth trend can be represented empirically 
by a function of the form $T=a\exp (bt+ct^2)$, where $T$ is the  
global surface temperature and $t$ is the time, with $a,b$ and $c$ free 
parameters. This is shown by the dotted curve in figure \ref{fig3}.  
The dash dotted curve shows this smooth trend with the long term CR ionization rate 
from figure \ref{fig2} added (inverted and arbitrarily normalised to give the 
best visual representation of the temperature data). The similarity between the 
deviations from the trend of the CR and temperature data 
is illustrated in more detail in figure \ref{fig3}(b) where the differences 
from the smooth trend are plotted against time. 

The similarity between the solar activity measured by the SSN and CR and the 
temperature deviations is striking. The temperature deviations seem 
to be in time with the SSN variations rather than delayed by 2-4 years 
as observed for the CR deviations. Hence, assuming that this is a real 
correlation between global surface temperature and solar activity,  
it is more probable that the deviations arise from the effects of a 
phenomenon, such as SI, rather than the effects of CR which are known to 
lag behind the SSN changes. 
%which travels directly 
%from the Sun to the Earth at the speed of light rather than propagating with 
%the speed of the solar wind to the outer reaches of the Solar System. 

\begin{figure}[ht]
\includegraphics[width=33pc]{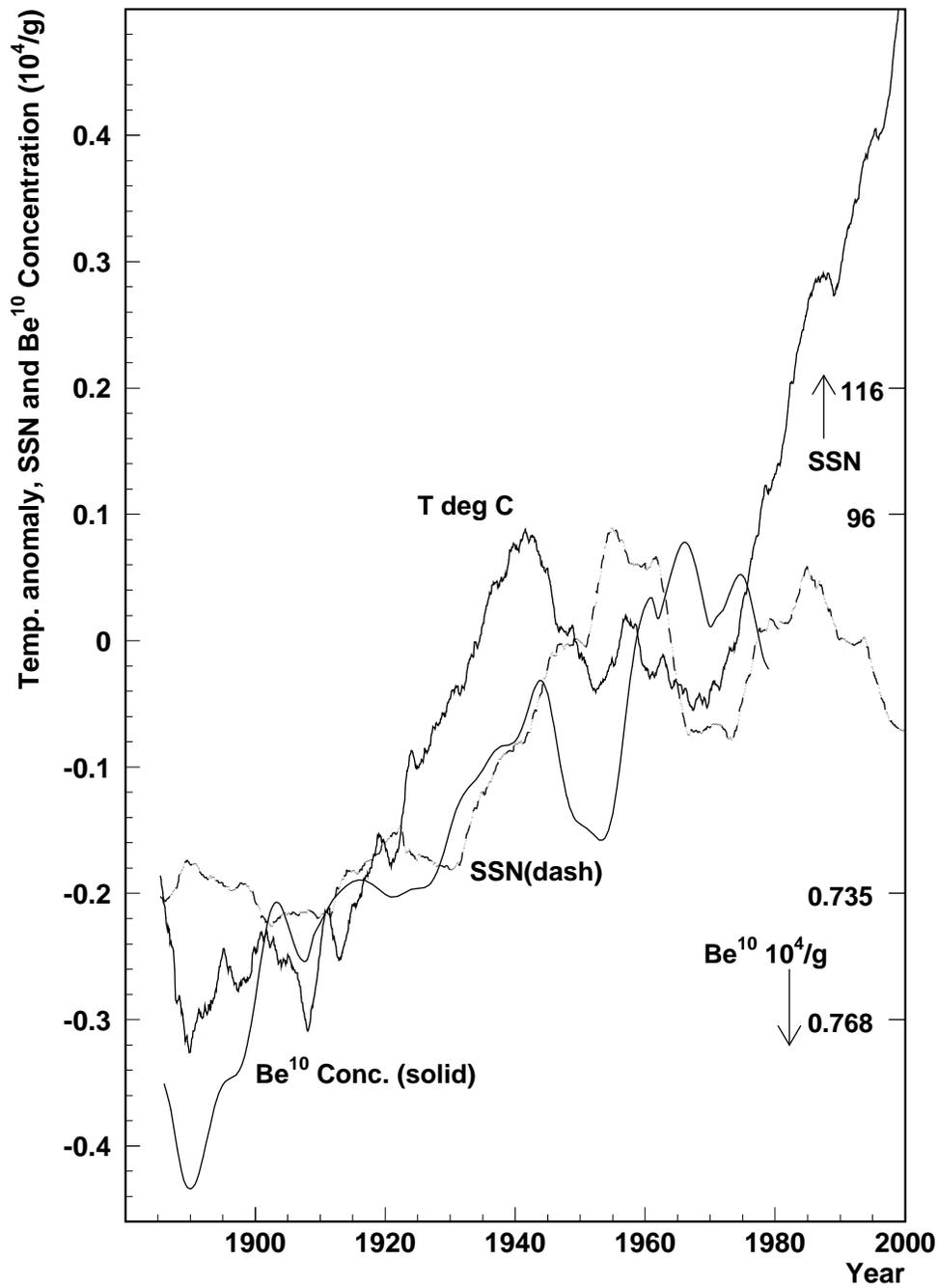}
%\hspace{2pc}%
%\vspace{-20pc}
%\begin{minipage}[b]{18pc}
\caption{\label{fig4p} The global surface temperature anomaly, 
the SSN (dashed curve) and the $^{10}$Be data from the Greenland ice 
core \cite{Beer1,Beer2} all with the 11 year smoothing. 
The scales for the SSN and $^{10}$Be concentration are shown on the right 
hand axes. Note that the $^{10}$Be scale is inverted for illustration 
i.e. it increases vertically downwards.}
%\end{minipage}
\end{figure}

Although the results of the simulations are only available for dates 
after 1951, it is instructive to examine the period before this using such 
data as are available. Figure \ref{fig4p} shows the temperature anomaly
compared to the SSN and  $^{10}$Be data \cite{Beer1,Beer2}, applying the 
same 11 year smoothing to each. Here we use the  $^{10}$Be data 
as a rough proxy for the CR ionization rate. It is 
apparent that the good correlation seen after 1956 shown in figure \ref{fig3} 
does not continue before this date. However, the $^{10}$Be data may not be 
a good proxy for the ionization rate in the troposphere since they show a 
different modulation due to the 11 year solar cycle than other CR measurements 
\cite{Nikitin}. In addition, they produce a different long term behaviour 
after 1956 than that shown in figure \ref{fig1} as can be seen from the overlap 
region in time in figures \ref{fig2} and \ref{fig4p}. 
This cosmogenic isotope is thought to be produced in the stratosphere by somewhat 
lower energy primaries than those responsible for the bulk of the ionization 
in the troposphere \cite{Webber}. Hence it could be expected to have different 
properties from the ionization in the troposphere. 
Thus, there is some possibility that the correlation seen after 1956 is real.

%In addition, it should be noted that the 
%peak in the temperature centred at 1940 is thought to arise from warming in the 
%early part of the century with aerosol contributions after 1940 then causing 
%some cooling \cite{Haigh}.
%Hence it is not surprising that such effects do not appear in the SSN and CR 
%data. Similarly, in \cite{Lean} the observed global mean temperature profile from 1880 
%to the present was well reproduced by a model involving corrections for volcanic aerosols, 
%the ENSO tropical temperature index, greenhouse gases and tropospheric aerosols,  
%none of which would be expected to be correlated to SSN or CR. 

\section{The effects of CR and SI on the global surface temperature}
%\section{The effects on the global surface temperature}

\subsection{Cosmic Rays}
The data in figure \ref{fig2} show that 
between the years 1956 to 2001 the global average CR rate has oscillated 
with the final value being roughly 1$\%$ higher than the starting value 
with an averaged fractional change of $0 \pm 0.2\%$. 
If global warming were being influenced by changes in ionization producing 
changes in low level cloud cover, as hypothesised 
in \cite{MandS,MandS2}, then the increase in the CR rate would be expected 
to produce a lowering of the globally averaged 
temperature of the Earth during this period rather than the increase shown in 
figure \ref{fig3}a. 
%. This would arise since an increase in the CR rate leads to 
%more ionization in the atmosphere, i.e. an increase in low level cloud cover 
%according to the hypothesis of \cite{Palle,MandS}, which would lead to a lowering 
%of the global surface temperature. 
This already is evidence against CR being a 
large contributor to global warming, as pointed out by Lockwood 
and Fr\"ohlich \cite{LandF}.  

In order to derive an upper limit on the global temperature rise which can 
be ascribed to changes in ionization from CR we adopt the hypothesis 
of \cite{MandS,MandS2} and assume that the oscillation 
in temperature shown in figure \ref{fig3} results from the oscillation in 
the ionization rate shown in figure \ref{fig2}.  
We observe in figure \ref{fig2} that the CR ionization rate  
oscillation is $\pm$1.5$\%$ which is anticorrelated with the temperature 
oscillation in figure \ref{fig3} of amplitude $\mp$0.07$^\circ$C. 
Any long term decrease in the CR ionization rate since 1956 is less than 
the amplitude of the cyclic variation. Hence, any long term change in the 
global temperature must be less than the amplitude of cyclic variation 
in the temperature i.e. less than 0.07$^\circ$C. This is to be compared with 
an observed rise of 0.5$^\circ$C (see figure \ref{fig3}). 
Thus, within our assumptions,  
less than 14$\%$ of the observed global warming since 1956 is attributable 
to the changing CR rate.    

\subsection{Solar Irradiance}

A similar argument can be applied to the case of SI. 
The data in figure \ref{fig2} show that 
between the years 1956 to 2001 the SI rate has oscillated 
about the mean with an amplitude of $\sim$0.1 Wm$^{-2}$. 
The final value is 0.09Wm$^{-2}$ higher than the starting value.  
From this it can be seen that the SI could have increased  
since 1956 by an amount which is less than the amplitude of the 
observed cyclic variation. If we attribute the oscillation in 
temperature to the cyclic change in SI, a change in SI of 
0.1 Wm$^{-2}$ causes a change in the global temperature 
of 0.07$^\circ$C. Since the change in total SI since 1956 
is less than 0.1 Wm$^{-2}$, it follows that the total change 
in mean global temperature due to SI must be less than 
0.07$^\circ$C. 

Hence,  within our assumptions, less than 14$\%$ of the observed 
global warming since 1956 is attributable to changes in SI.  

\section{Conclusions}

The long term variation of the cosmic ray ionization rate  
has been studied. This rate shows a   
cyclic variation with a period of roughly twice the 11 year cycle 
for the data available since the 1950s.  
The structures seen in the variation of the long 
term cosmic ray ionization rate with time are shown to be present in the 
variation of the mean daily sun spot number, solar irradiance and in 
the variation of the mean global surface temperature. Hence we report 
a possible observation of a cyclical variation in each of these quantities 
of a similar period.  
The cyclic variation of the global temperature is found to be in 
phase with the solar cycle as measured from the sun spot numbers 
and the solar irradiance and in antiphase with the cosmic ray variation.  
However, the cyclic variation of the CR cycle is delayed by 
2-4 years. This indicates that, if it is real, the correlation is most  
likely caused by direct solar activity rather than by cosmic rays. 

The long term variations of each of the cosmic ray rate and the solar 
irradiance are observed to be less than their cyclic variations. 
Therefore, assuming that there is a causal link between either 
of them with 
the mean global surface temperature, the long term variation of the 
temperature must be less than the amplitude of its cyclic variation 
of 0.07$^\circ$C. Hence  within our assumptions, 
the effect of varying solar activity, either 
by direct solar irradiance or by varying cosmic ray rates, must be less 
than 0.07$^\circ$C since 1956 i.e. less than 14$\%$ of the observed global 
warming. 

%Nevertheless assuming that the correlation is caused by cosmic rays it 
%is possible to use it to estimate the contribution of the 
%changing cosmic ray rate to global warming.  
%The global mean ionization rate has been observed to increase by 1$\%$ since 
%1956. 
%Postulating that the observed long term cosmic ray rate varies cyclically 
%with an overall downward trend during this time, a decrease of the rate of 
%ionization due to cosmic rays of 0.7$\%$ per 25 years is possible.
%  According to the hypothesised link between cosmic rays and cloud cover, 
%the observed increased rate of ionization over the last half century 
%should have led to global cooling 
%rather than the observed global warming. However, with the postulated 
%downward trend, global warming at the level of 18$\%$ of that observed 
%becomes possible. We conclude therefore, based on these assumptions, 
%that $\lesssim$18$\%$ of the 
%global warming observed between 1975-2000 can be attributed to changes of 
%ionization from cosmic rays.  

\section{Acknowledgements}
We are grateful to G. Bazilevskaya, J. Beer, J. Haigh and I.G. Usoskin 
for valuable discussions and for  
providing us with computer readable tables of data on the ionization rate 
from the Geiger counter measurements, the Dye 3 ice core data, the SI data 
and the ionization simulations, respectively.  
We are also grateful to our anonymous referees for their comments which have 
helped us to improve the paper. We thank the Dr. John C. Taylor 
Charitable Foundation for financial support.

\newpage

%figure 1

%\newpage

%figure 2

%figure 3

%figure 4p

%\end{minipage}

\end{document}